\documentclass[a4paper,fleqn,usenatbib]{mnras}

\usepackage[T1]{fontenc}
\usepackage{ae,aecompl}

\usepackage{caption}
\usepackage{natbib}
\usepackage{graphicx}
\usepackage{textcomp}
\usepackage{latexsym}
\usepackage{varioref}
\usepackage{xspace}
\usepackage{makeidx}
\usepackage{verbatim}
\usepackage{tabularx}
\usepackage{epstopdf}
\usepackage{amsmath}
\usepackage{array}
\usepackage{rotating}
\usepackage{hyperref}
\usepackage{float}
\usepackage{amssymb}

\def\apj{ApJ}
\def\apjs{ApJS}
\def\apjl{ApJL}
\def\aj{AJ}
\def\mnras{MNRAS}

\def\aap{A\&A}

\def\Kepler{\textit{Kepler}}

\def\K2{\textit{K2}}

\def\EPIC{EPIC 204278916}
\def\KIC{KIC 8462852}

\title[Peculiar dips in EPIC 204278916]
{The peculiar dipping events in the disk-bearing young-stellar object EPIC 204278916}
\author[S. Scaringi et al.]
{S.~Scaringi$^{1}$\thanks{E-mail: simo@mpe.mpg.de}, C.~F.~Manara$^{2}$, S.~A.~Barenfeld$^{3}$, P.~J.~Groot$^{4}$, A.~Isella$^{5}$, \newauthor M.~A.~Kenworthy$^{6}$, C.~Knigge$^{7}$, T.~J.~Maccarone$^{8}$, L.~Ricci$^{9}$, M.~Ansdell$^{10}$\\ 
$^{1}$Max-Planck-Institute f{\"u}r Extraterrestriche Physik, D-85748 Garching, Germany\\
$^{2}$Scientific Support Office, Directorate of Science, European Space Research and Technology Centre (ESA/ESTEC), Keplerlaan 1,\\ 2201 AZ Noordwijk, The Netherlands\\
$^{3}$California Institute of Technology, Department of Astronomy, MC 249-17, Pasadena, CA 91125\\
$^{4}$Department of Astrophysics/IMAPP, Radboud University Nijmegen, P.O. Box 9010, 6500 GL Nijmegen, The Netherlands\\
$^{5}$Department of Physics and Astronomy, Rice University, 6100 Main St. Houston, TX 77005, USA\\
$^{6}$Leiden Observatory, Leiden University, PO Box 9513, NL-2300 RA Leiden, the Netherlands\\
$^{7}$School of Physics and Astronomy, University of Southampton, Hampshire SO17 1BJ, United Kingdom\\
$^{8}$Department of Physics and Astronomy, Texas Tech University, Box 41051, Lubbock, TX 79409-1051, USA\\
$^{9}$Harvard-Smithsonian Center for Astrophysics, 60 Garden Street, Cambridge, MA 02138, USA\\
$^{10}$Institute for Astronomy, University of Hawai`i at M\={a}noa, Honolulu, HI, USA}

\date{Accepted XXX. Received YYY; in original form ZZZ}

\pubyear{2016}

\begin{document}
\label{firstpage}
\pagerange{\pageref{firstpage}--\pageref{lastpage}}
\maketitle

\begin{abstract}
\EPIC\ has been serendipitously discovered from its \K2\ light curve
which displays irregular dimmings of up to $65\%$ for $\approx$25
consecutive days out of 78.8 days of observations. For the remaining
duration of the observations, the variability is highly periodic and
attributed to stellar rotation. The star is a young, low-mass (M-type)
pre-main-sequence star with clear evidence of a resolved tilted disk
from ALMA observations. We examine the \K2\ light curve in detail and
hypothesise that the irregular dimmings are caused by either a warped
inner-disk edge or transiting cometary-like objects in either circular
or eccentric orbits. The explanations discussed here are particularly
relevant for other recently discovered young objects with similar
absorption dips.
\end{abstract}


%%%%%%%%%%%%%%%%%%%%%%%%%%%%%%%%%%%%%%%%%%%%%%%%%%

%%%%%%%%%%%%%%%%% BODY OF PAPER %%%%%%%%%%%%%%%%%%

\section{Introduction}

Studies of light curves of stars are a proxy for several physical processes, both
at the stellar surface or in their surroundings. Periodic variability of
the observed stellar flux has long been used to measure stellar
rotation periods (e.g. \citealt{stassun99}). Variability in young
stellar objects (YSOs) is also related to the presence of a
protoplanetary disk. Material in the disk can obscure the central star
(e.g. \citealt{Herbst94,Alencar10}), and the highly variable accretion
onto the central star modifies the emission from the system (e.g.
\citealt{bertout88,bouvier07}). The variability of YSOs is known to have
different degrees of periodicity and flux symmetry which are possibly
related to different processes (\citealt{cody14}). Recently, \textit{CoRoT}
and \Kepler/\K2\ observations of main sequence stars and YSOs have shown
deep and short dips in light curves that can be explained by the presence of a family of comets orbiting around the stars
(\citealt{boyajian15, bodman15}) or more generally transiting circumstellar material (\citealt{ansdell16a}), or through occulting
material at the inner edges of a circumstellar disk (\citealt{mcginnis15,ansdell16a}). The
latter is possible if the inner disk is observed almost edge-on, and offers a
unique opportunity to study the properties of dust and gas in the inner
region of protoplanetary disks. If the dimming events are caused by
transiting circumstellar clumps originating in the disk, these can be used to constrain the size of
planetesimals in the disk, which is a necessary step for planet formation
but difficult to observationally detect \cite{testi14}.

More recently, \cite{ansdell16b} have found 3 YSO dippers with a wide range of inclination angles, demonstrating how edge-on disks are not a defining characteristic of dipping YSOs. At this stage it is important to study a variety of dipper YSOs in detail and explore possible scenarios to explain their behaviour and their relevance for planet formation studies.

We have serendipitously discovered a YSO dipper star observed with \K2,
giving us the possibility to further investigate these peculiar systems.
The target discussed here (\EPIC, 2MASS J16020757$-$2257467) is a young
star located in the Upper Scorpius sub-group of the Scorpius-Centaurus
OB association (\citealt{ScoCen}) with very high membership probability
based on proper motion studies (99\%, \citealt{bouy09}). This region has
a mean age of $\sim$5 Myr (\citealt{preibisch02}, with an upper limit of
$\sim$11 Myr; \citealt{pecaut12}). Our target is a single star
(\citealt{kraus07}) of spectral type M1 and has a logarithmic bolometric
stellar luminosity log$L_\star/L_\odot$ = 0.15 (\citealt{preibisch02}).
The inferred radius of this star is $R_\star$ = 0.97 $R_\odot$, while
the stellar mass is $M_\star \sim$ 0.5 $M_\odot$, depending on the
evolutionary models used (\citealt{baraffe98,siess00}). 

At this age, the majority of young stars have already dispersed their
circumstellar disk and are not accreting material from the disk anymore
(e.g., \citealt{fedele10}). Spectroscopic observations of the H$\alpha$
line of this target report a very small equivalent width
(EW = -3.2 \AA, \citealt{preibisch02}), consistent with very little or
no accretion. However, this target shows an infrared excess due to the
presence of a protoplanetary disk in WISE data (\citealt{luhman12}).

The next section will introduce the \K2\ and ALMA data and analysis
procedure to obtain the reduced light curve (with corresponding Fourier
transform) and the ALMA intensity map. Section \ref{sec:results}
discusses the observations in the context of various interpretations,
including a protostellar disk origin and cometary-like transits in
either circular or eccentric orbits. We discuss our results in section \ref{sec:discussion} and give our conclusions in Section \ref{sec:conclusion} with future prospects for determining the real nature of \EPIC.

\section{Data reduction and analysis}\label{sec:obs}

\subsection{\textit{K2 light curve}}

\EPIC\ was observed by the \K2\ mission (\citealt{borucki}) during
Campaign 2 between August 23 and November 13 2014 (78.8 days) and has a
registered \Kepler\ magnitude ($Kp$) of $13.8$ in the \K2\ Ecliptic
Plane Input Catalog (EPIC). Here we analyse long cadence (LC, 29.4
minutes) data obtained from the Mikulski Archive for Space Telescope
(MAST) archive\footnote{\url{http://archive.stsci.edu/k2/}}. The data is
provided in raw format, consisting of target pixel data. For each 29.4
minute exposure we thus have a 12 $\times$ 11 pixel image centred on the
target. 

\begin{figure*}
\includegraphics[width=1\textwidth]{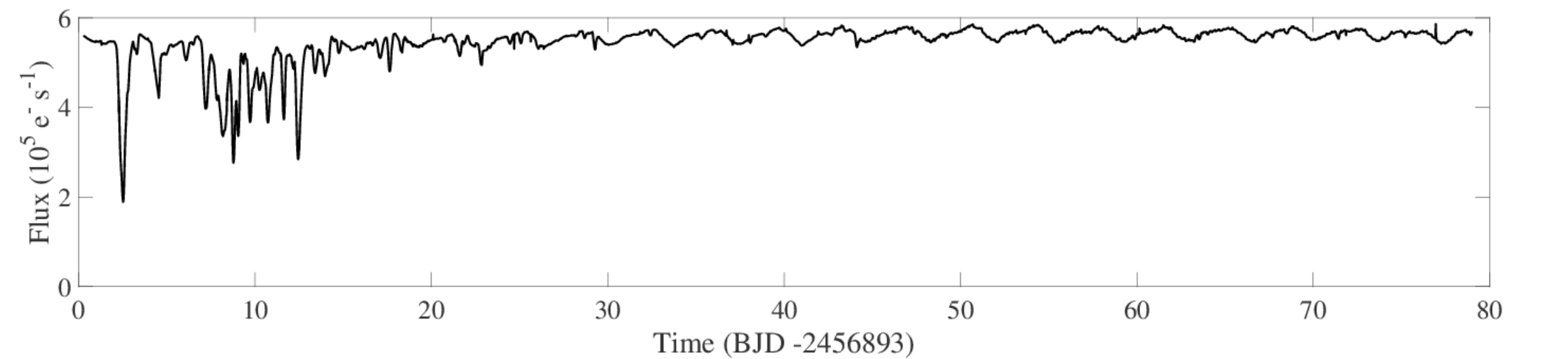}
\includegraphics[width=1\textwidth]{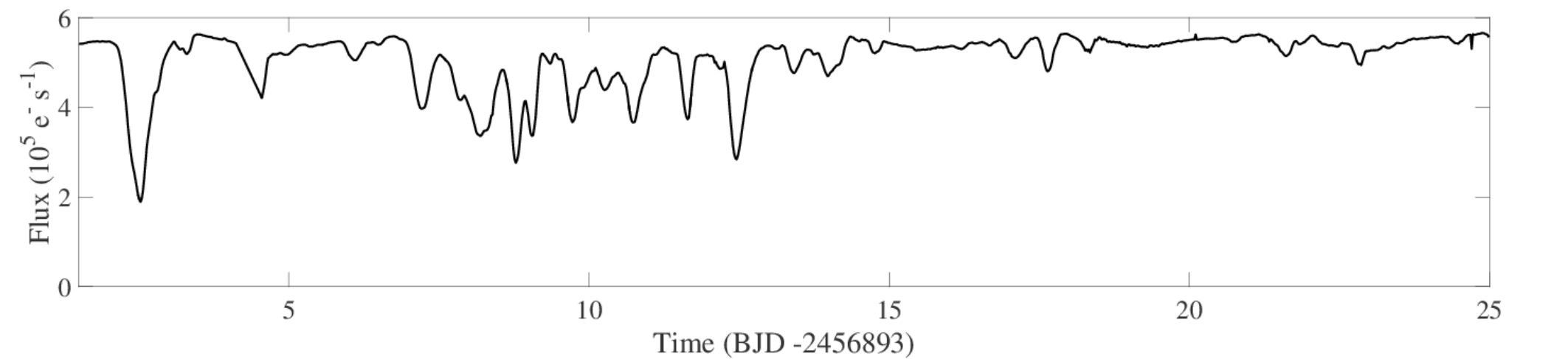}
\caption{\EPIC\ light curve. The system was observed for over 78.8 days at 29.4 minute cadence. The units on the y-axis are electrons/second, and can be converted to \Kepler\ magnitudes $Kp$ using the conversion found in the \Kepler\ Instrument Handbook. The top panel shows the full light curve, whilst the bottom panel zooms into the first 25 days of observation where the dipping events are observed.}
\label{fig:LC}
\end{figure*}

As no other sources were present within the 12 $\times$ 11 pixel images,
we create the light curve by manually defining a large target mask as
well as a background mask. A large target mask is required due to
occasional small scale jittering of the spacecraft, resulting in the
target moving slightly from its nominal position. The dataset consists
of 3856 individual target images. From these we remove 93 observations
because of bad quality due to occasional spacecraft rolls or due to
cosmic rays. We produce the light curve by summing together all target
pixels for each exposure, and subtract the average background obtained
from the background pixel mask. The obtained light curve is shown in
Fig. \ref{fig:LC}. The same procedure has been used to extract other
\K2\ light curves of isolated sources
(\citealt{scaringi15a,scaringi15b}).

\begin{figure*}
\includegraphics[width=1\textwidth]{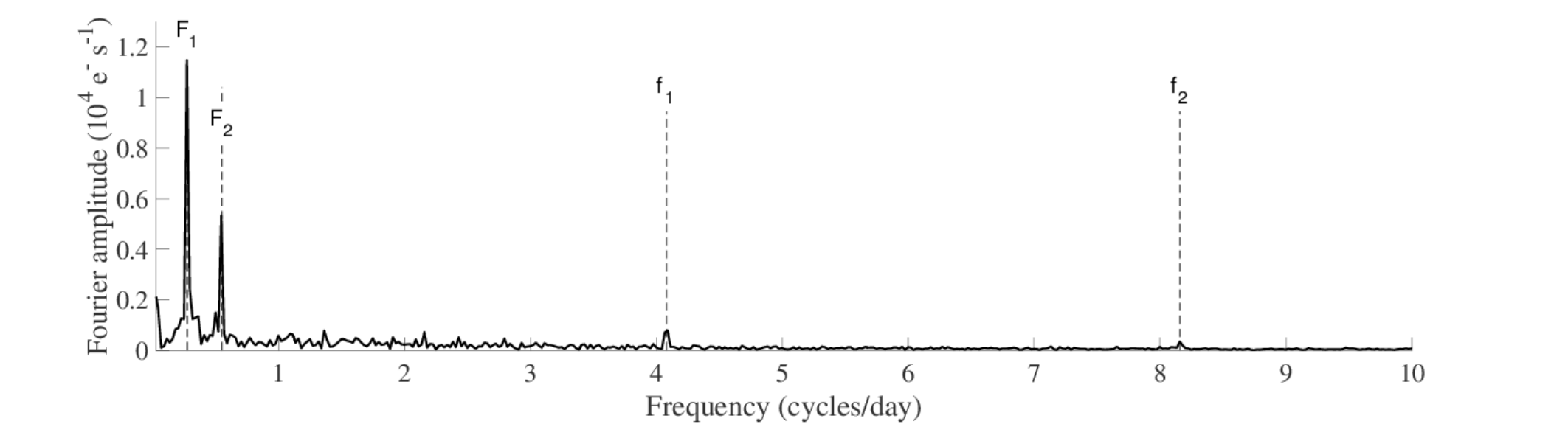}
\caption{Fourier transform of LC \K2\ data for \EPIC\ starting from day
25. A peak at $F_1 = 0.2743$ cycles/day (with corresponding harmonic
$F_2$) is visible and at $f_1=4.080$ cycles/day (with corresponding
harmonics $f_2$, $f_3$, $f_4$) and $f_5$. We associate $F_1$ to the
possible stellar rotation, whilst $f_1$ can be attributed to spacecraft attitude adjustments.}
\label{fig:FFT}
\end{figure*}

\begin{figure}
\includegraphics[width=0.48\textwidth]{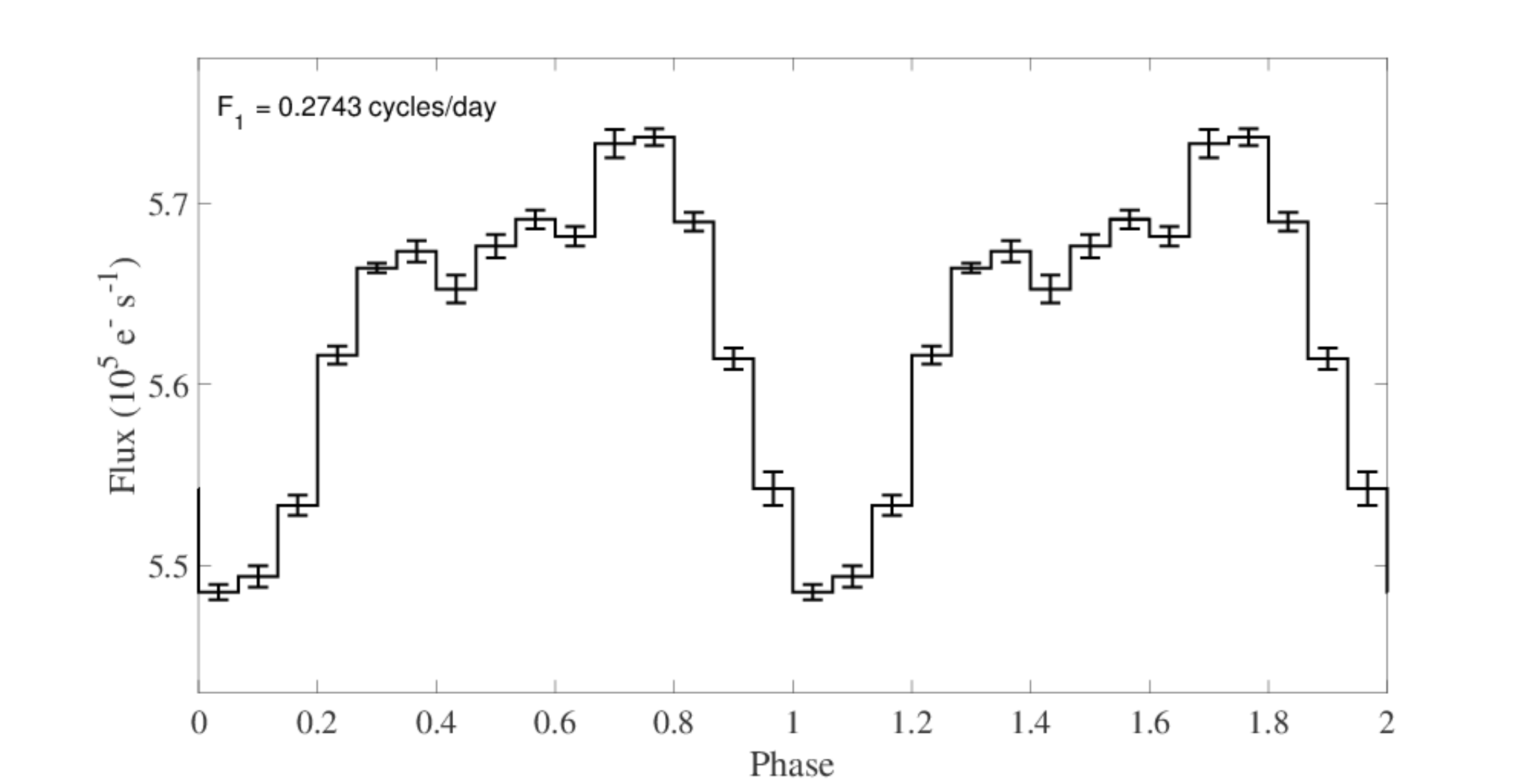}
\caption{\EPIC\ light curve folded on the stellar rotation period of 3.646 days.}
\label{fig:folded}
\end{figure}

To locate any significant periodicities we carry out a Fourier transform
of the light curve excluding the first 30 days (in order to not be
affected by the initial dipping period, see Fig. \ref{fig:LC}). Fig.
\ref{fig:FFT} shows our result, revealing two distinct periodicities of
3.646 days ($F_1 = 0.2743$ cycles/day) and 0.245 days ($f_1 = 4.080$
cycles/day), together with corresponding harmonic frequencies. We
associate the 3.646-day signal to stellar rotation, as this
is a typical rotational period for young stars stars observed by \Kepler\
(\citealt{ansdell16a,vasco15}), and show the phase-folded profile in Fig.
\ref{fig:folded}. The 0.245-day signal is associated with the spacecraft
thruster firing to re-adjust attitude every $\sim 6$ hours.

To study the dipping behaviour in more detail we have removed the
3.646-day periodicity from the full light curve. To do this we first
determined an accurate ephemeris of

\begin{equation}
\text{BJD}_{\text{min}} = 2 456 955.7082(18) + 3.646221(53) \times N ,
\label{eq:eph}
\end{equation}

\noindent where $N$ is the cycle number and the ephemeris are given in Barycentric Julian Date (BJD) at minimum light.

\begin{figure*}
\includegraphics[width=1\textwidth]{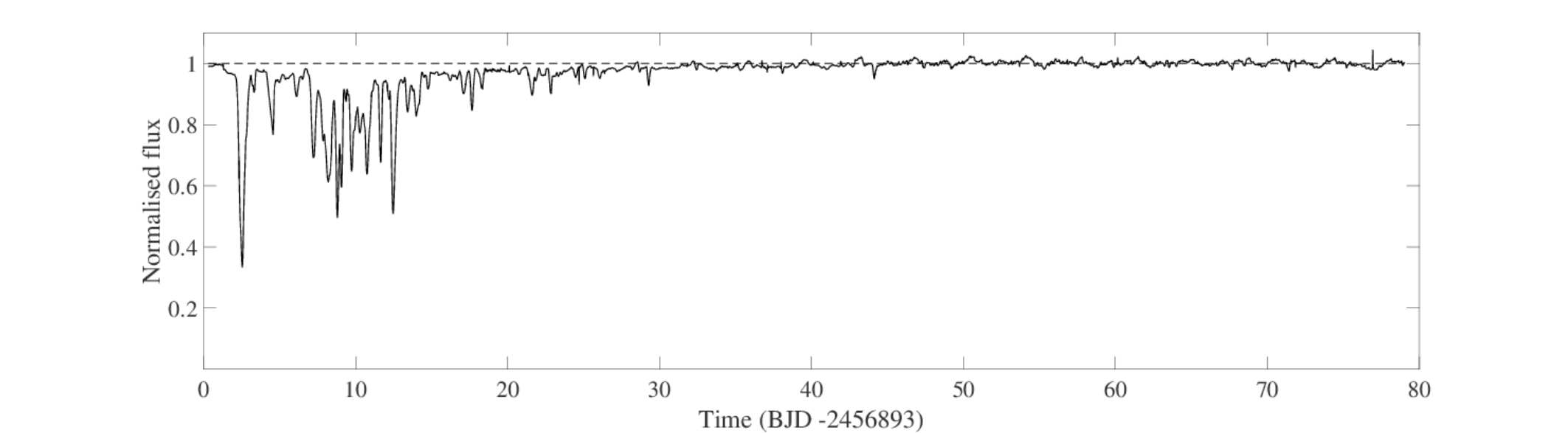}
\caption{Normalised \EPIC\ light curve with the 3.646-day periodicity removed.}
\label{fig:LC-Spin}
\end{figure*}

We then expand and replicate the phase folded light curve shown in Fig.
\ref{fig:folded} to match the full light curve and interpolate it on the
same temporal grid. Fig. \ref{fig:LC-Spin} shows the full light curve
with the 3.646-day periodicity removed. The periodicity removal
is not perfect (probably due to small changes in the pulse profile over
time) but shows a significant improvement over the original light curve.
We use this curve in subsequent analysis.

\subsection{ALMA observations}

\begin{figure}
\includegraphics[width=0.45\textwidth]{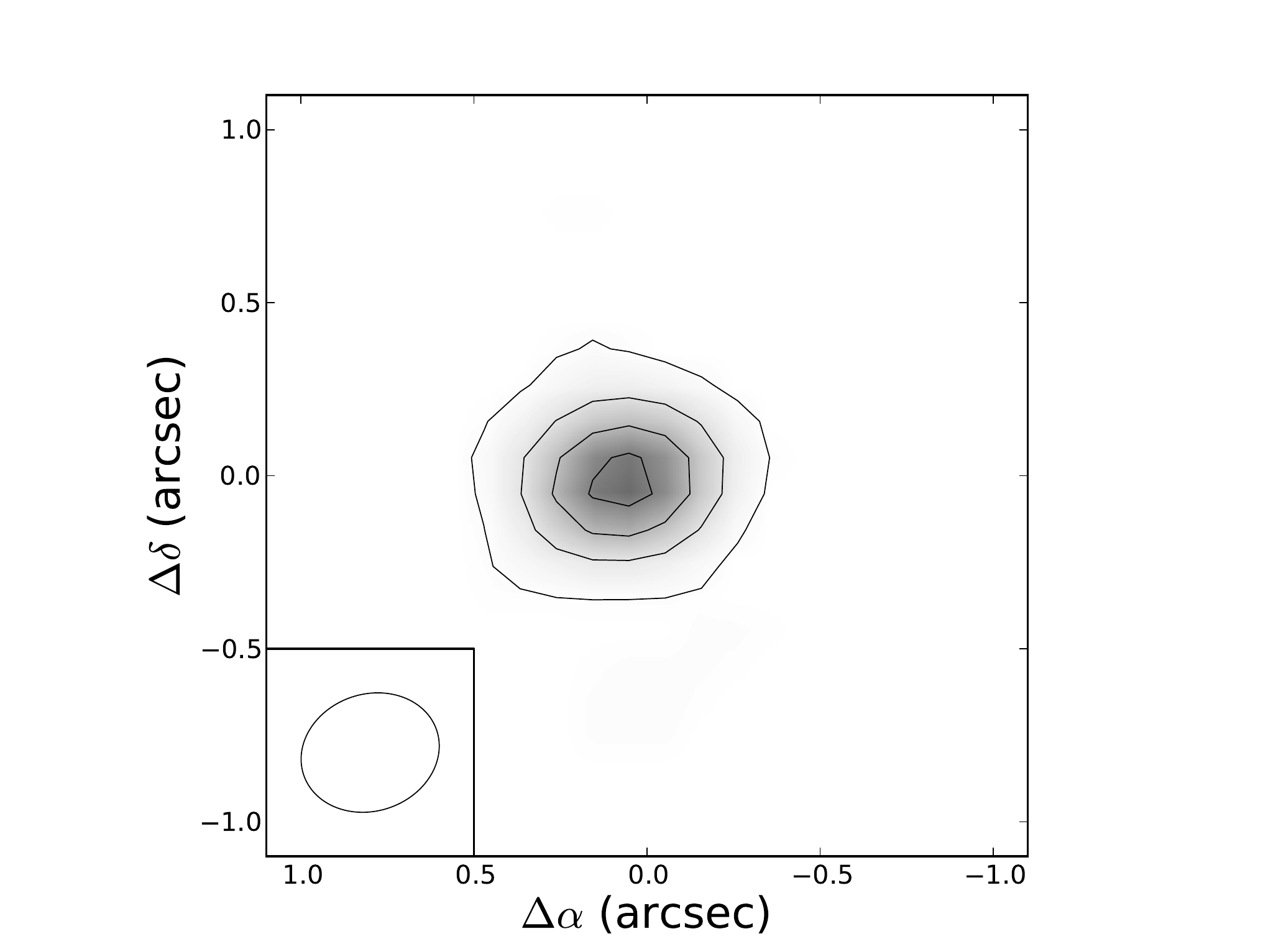}
\caption{ALMA continuum intensity map for \EPIC. Isocontours for 3-,10-,17- and 23-$\sigma$ are overlaid on the image. The ALMA beam size is shown in the bottom-left corner.}
\label{fig:alma}
\end{figure}

ALMA Cycle 2 observations of \EPIC\ were obtained on 2014 June 30 and
2014 July 7 (UT) in four spectral windows centred at 334.2, 336.1,
346.2, and 348.1 GHz for a mean frequency of 341.1 GHz (0.88 mm). The
bandwidth of each window is 1.875 GHz. Thirty six antennas were included
in the array, with baselines ranging from 16-650 m for an angular
resolution of 0.34 arcseconds. The ALMA data were calibrated using the
CASA package. The reduction scripts were kindly provided by NRAO. These
include atmospheric calibration using the 183 GHz water vapour
radiometers, bandpass calibration, flux calibration, and gain
calibration. For further details on the observations and data reduction,
see \cite{barenfeld16}. The protoplanetary disk in \EPIC\ is resolved
with ALMA at high signal-to-noise, both in the continuum (see Fig. \ref{fig:alma}) and in the
CO(J=3$-$2) line (\citealt{barenfeld16}). An elliptical Gaussian fit of
the surface brightness profile to the continuum visibility data results
in an axis ratio of 0.55 $\pm$ 0.13, implying a disk inclination of 57
$\pm$ 9 degrees and assuming a circular disk.

\section{Results} \label{sec:results}

In this section we will go through possible scenarios to explain the observed large amplitude dipping events in \EPIC. 

\subsection{Proto-stellar disk origin} \label{sec:disk}

Provided with a favourable disk inclination, the observed dips in the \EPIC\
light curve could be caused by non-axisymmetric structures in the inner
disk edge occulting the star. Because of the large observed dips
($\approx65\%$), the occulting material must have  a large scale height
comparable to the size of the star. If the disk warps producing the
occultations are orbiting the star with a Keplerian period equal to the
stellar rotation, we can infer a maximum co-inclination
(\citealt{ansdell16a}) in degrees

\begin{equation}
i_{max} \lesssim 27 \frac{R_{\ast}}{R_{\odot}} \frac{P_{rot}}{\text{days}}^{-2/3} \frac{M_{\ast}}{M_{\odot}}^{-1/3} .
\label{eq:}
\end{equation}

\noindent Adopting $P_{rot}=3.646$ days, $R_{\ast}=0.97R_{\odot}$ and $M_{\ast} = 0.5M_{\odot}$ we obtain a maximum inclination $i_{max}=14$ degrees. This is somewhat in contrast with the $3\sigma$ level for the disk inclination of $i=57$ degrees obtained from the ALMA image.

Other YSOs have been observed to display non-periodic dipping events
(see e.g. \citealt{sousa16,ansdell16a,ansdell16b}), and it is
possible \EPIC\ is also part of this ``dipper'' class of YSOs. In
particular, \EPIC\ supports the idea that nearly edge-on viewing
geometries are not a defining characteristic of dippers. This in turn
motivates the exploration of alternative models. More specifically, the
clustering of large-amplitude dipping events such as those observed in
the \K2\ \EPIC\ light curve have not been previously observed in other
systems. The only other YSO dipper which appears to resemble \EPIC\ as observed with \K2\ is EPIC 204530045 (see Fig.~1 of \citealt{bodman16}). However, on close inspection, it appears that the dipping behaviour between \EPIC\ and EPIC 204530045 differs in that a) the dips in \EPIC\ always appear to return to the stellar flux whilst there is a clear downward general trend for EPIC 204530045, b) whilst \EPIC\ solely displays dipping, the lightcurve of EPIC 204530045 more resembles flickering and c) the dip depth is nearly a factor 3 larger for \EPIC\ when compared to EPIC 204530045. All three arguments seem to suggest that whilst the \K2\ lightcurves of both \EPIC\ and EPIC 204530045 appear similar, the phenomenology observed in EPIC 204530045 seems to be accretion driven (change in instantaneous mass transfer rate on the stellar surface), whilst \EPIC\ has all the characteristics of being ``transited''. Thus, if \EPIC\ is also part of the YSO ``dipper'' class as those discussed in \cite{bodman16}, then it constitutes a somewhat special case. 

\subsection{Hill sphere ring system}

\begin{figure*}
\includegraphics[width=1\textwidth]{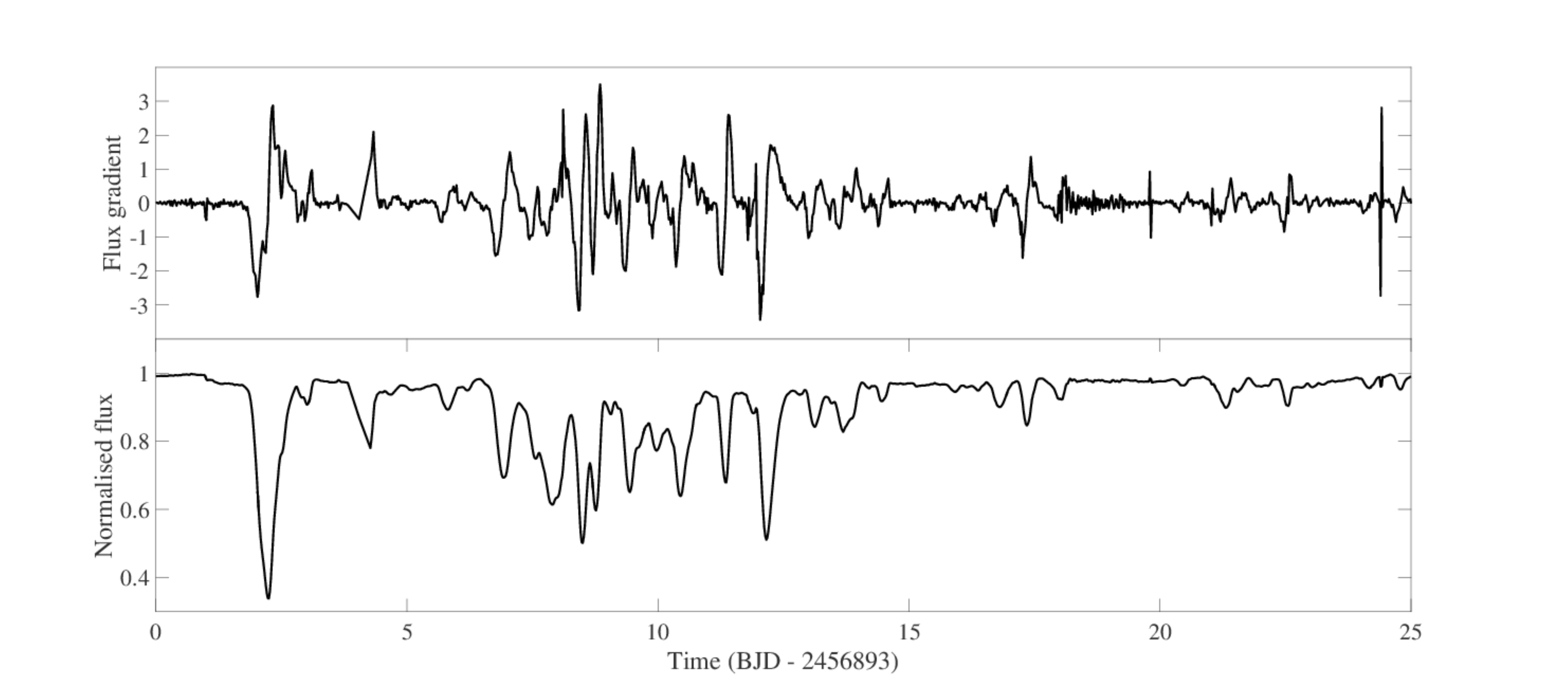}
\caption{Bottom-panel: Normalised \EPIC\ light curve (with the 3.646-day periodicity removed) for the first 25 days of observation. Top-panel: Corresponding lightcurve gradient.}
\label{fig:fluxGradient}
\end{figure*}

The rapid fluctuations over the first 25 days of observations are similar to those seen towards
J1407, another young star in the Sco-Cen association. In May 2007 a
series of rapid fluctuations was seen over a 56 day period
(\citealt{mamajek12}) and interpreted as a giant ring system filling the
Hill sphere of the unseen secondary companion
(\citealt{kenworthy15a,kenworthy15b}). The series of dips after the first
initial deep dip show hints of being symmetric in time, around $T=10$
days in Fig. \ref{fig:fluxGradient} (bottom panel), consistent with a ring system that is inclined with
respect to our line of sight. The gradient of the light curve of such an
inclined ring system will show small gradients near the point of closest
projected approach as the star moves parallel to ring edges, and steeper
light curve gradients at other times. This is not seen in the gradient
of the light curve as a function of time (Fig. \ref{fig:fluxGradient}, top-panel), and so this is considered an
unlikely hypothesis for these dips. 

\subsection{Transiting circumstellar clumps} \label{sec:comets}

It is possible that the dips in some YSOs might be caused by transiting
circumstellar objects, similar to what has been proposed for \KIC\
(\citealt{boyajian15,bodman15,marengo15}). The discussion
presented below is tailored to the dipping YSO \EPIC, but can be
extended to other dipping YSOs such as those discussed in
\cite{ansdell16b}.

\subsubsection{Transiting material in circular orbit}

\begin{figure}
\includegraphics[width=0.48\textwidth]{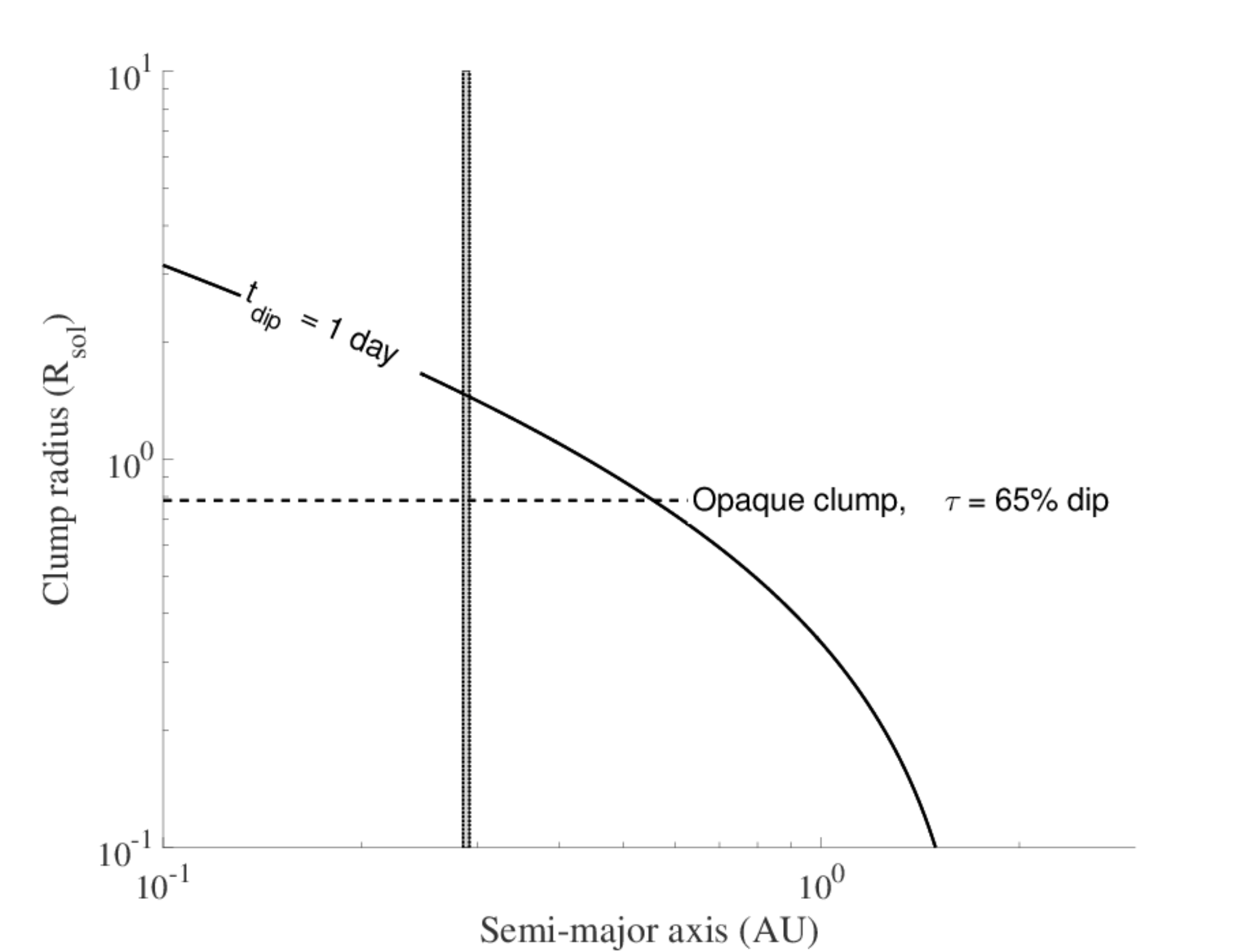}
\caption{Assuming 1-day dips, the figure shows the possible semi-major axis vs. clump radius values with the solid black line. The minimum clump size, obtained through the observed dip amplitude, is shown with the dashed line. The two vertical dotted lines show the minimum and maximum semi-major axis obtained with the non-periodicity and observed minimum velocity (for optically thick material) constraint respectively.}
\label{fig:Rclump}
\end{figure}

With the assumption that the transiting object(s) are in circular orbits
we can place tight constraints on the orbital parameters in
a plane defined by the semi-major axis ($a$) and the transiting clump
radius ($R_{c}$). These are included in Fig. \ref{fig:Rclump}, and
described in detail below.

\textit{Dip depth:}  Constraints on the clump size can be obtained from
the dip depth $\tau$, defined as 1 minus the normalised absorbed flux
during the dipping event (see Fig.\ref{fig:LC-Spin}). In the most
extreme scenario where the eclipsing clump is completely opaque,
max$(\tau) = (R_c/R_{\ast})^2$. The deepest event observed in the \EPIC\
light curve corresponds to $\tau\approx65\%$, thus implying that at
least some of the clumps are a sizable fraction of the parent star. This
is however a strong lower limit since a completely opaque spherical
clump would produce a symmetrical dip, which we do not observe in the
light curve (see Fig. \ref{fig:LC}).

\textit{Dip duration:} A transiting object will have a transverse velocity along the stellar equator of
\begin{equation}
v_t = \frac{2(R_c+R_{\ast})}{t_{dip}},
\label{eq:1}
\end{equation}

\noindent where $t_{dip}$ is the transit duration. If the object is on a circular orbit around a star of mass $M_{\ast}$ with semi-major axis $a$, then we can estimate the size of the transiting object with
\begin{equation}
R_c = \frac{t_{dip}}{2} \left( \frac{GM_{\ast}}{a} \right)^{1/2} - R_{\ast}.
\label{eq:2}
\end{equation}

\noindent Thus, the observed dip durations in the \EPIC\ light curve
can provide an estimate for the transiting clump size for circular
orbits. The longest dip duration of $t_{dip}=1$ day provides the most
stringent constraint, and is shown as a solid black line in Fig.
\ref{fig:Rclump}.

\textit{Light-curve gradient:} An outer constraint on the semi-major
axis can be derived using the largest gradient observed during a transit
event. Transiting material will change the light curve most rapidly when
it is optically thick and transiting through the stellar equator.
\cite{vanWerk14} provides the equation required to translate the
observed gradients into a minimum velocity ($v_{min}$) for transiting
material using the so-called ``knife edge'' model. Assuming then that
the material is in a circular orbit and optically thick the obtained minimum velocity
constraint of $39$km/s translates to a maximum semi-major axis of 0.2916
AU through

\begin{equation}
a_{max} = \frac{GM_{\ast}}{v_{min}^2} .
\label{eq:3}
\end{equation}

\textit{Non-periodicity:} Given that we do not observe a repetition
of the dipping events within the total light curve, we can place a constraint
on the orbital period to be longer than 78.8 days. This in turn sets a
lower limit on the semi-major axis of 0.2855 AU for circular orbits.

All constraints for circular orbits are displayed in Fig.
\ref{fig:Rclump}. The very tight semi-major axis constraints raise some
doubts on this interpretation, since they imply that the orbital period
of the transiting material is just slightly longer than the 78.7 day
\K2\ observations. Furthermore it is possible that the dipping events began before the start of the \K2\ observations, in which case circular orbits would be fully ruled out. It is also important to note that the resulting size for the transiting clump would be very large at $R_{c}\approx1.5R_{sol}$. 

\subsubsection{Transiting material in eccentric orbit}

ALMA observations have revealed the inclination of the circumstellar
disk  to be 57
$\pm$ 9 degrees. It is thus plausible that the
transiting material orbits out of the disk plane, in which case its
motion does not need to be constrained to a circular orbit. Introducing
an eccentric orbit is far less constraining than assuming
circular orbits. We can nevertheless use the observed minimum velocity (obtained through the light curve
gradient) and the minimum semi-major axis (obtained through the
non-periodicity of the dips) to provide some constraints.

For eccentric orbits with eccentricity $e$, the orbital velocity at pericentre and apocentre respectively are defined as

\begin{equation}
v_{per} = \sqrt{GM_{\ast} \frac{1+e}{a(1-e)}}
\end{equation}
\noindent and
\begin{equation}
v_{apo} = \sqrt{GM_{\ast} \frac{1-e}{a(1+e)}} .
\end{equation}

\begin{figure}
\includegraphics[width=0.48\textwidth]{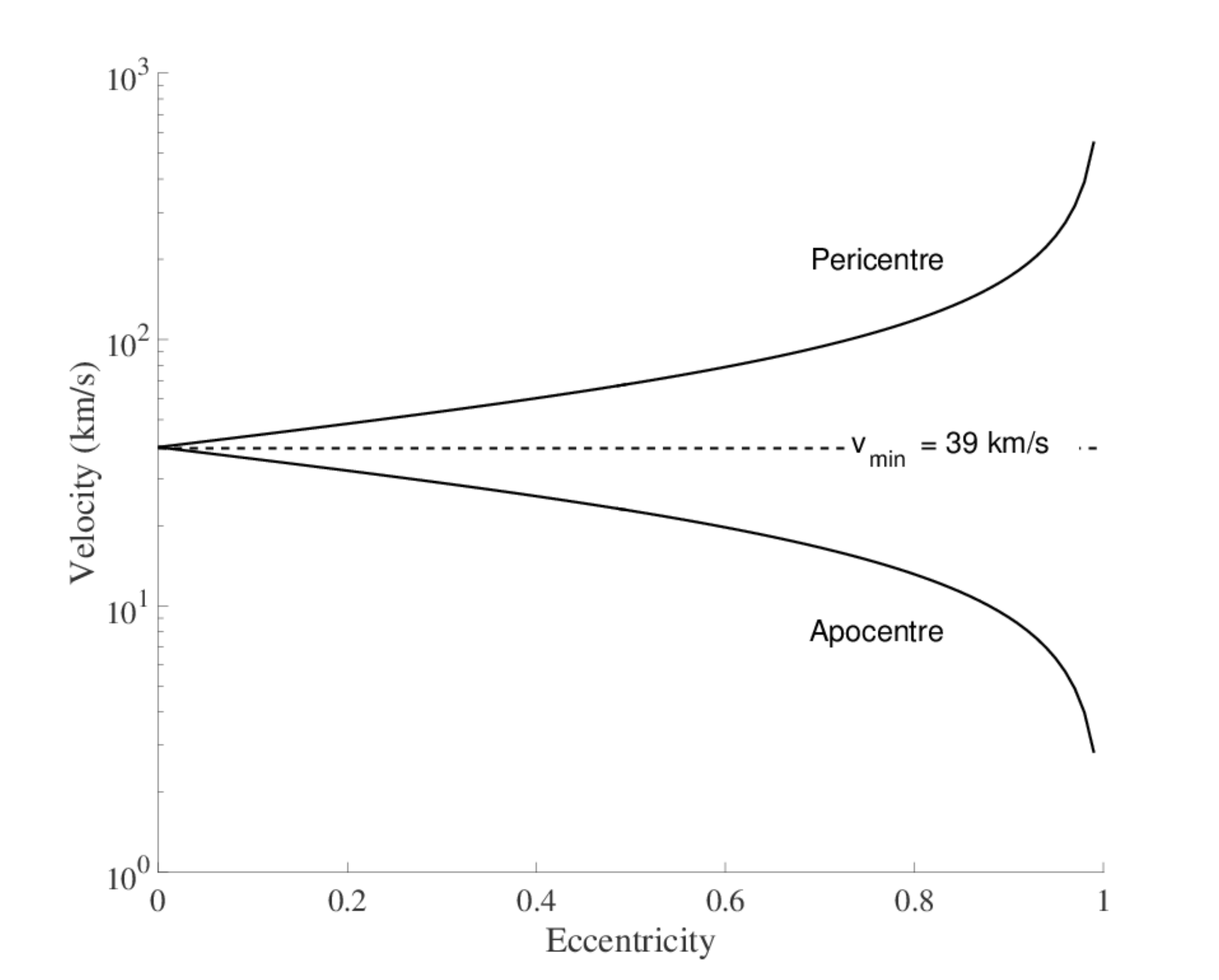}
\caption{Pericentre and apocentre orbital velocities for varying eccentricities with a constant semi-major axis of $a=0.2855$ (solid black lines). The dashed horizontal line marks the observed minim velocity of 39 km/s. If the observed dips in \EPIC\ are caused by transiting material on an eccentric orbit, then the observed transit must occur away from apocentre.}
\label{fig:orbVel}
\end{figure}

\noindent Fig. \ref{fig:orbVel} shows these velocities as a function of
eccentricity adopting the derived minimum semi-major axis of $a_{min} =
0.2855$ AU. The obtained velocities represent maximum velocities at both
pericentre and apocentre, since increasing the semi-major axis will
result in lower velocities. Plotted in Fig. \ref{fig:orbVel} is
the observed minimum velocity of $v_{min}=39$km/s obtained from the
light curve gradient of the dips. This already shows that if the orbit
is eccentric, then we are most likely observing the transit away from
apocentre. 

We can further expand on this analysis and speculate what the eccentricity would be if the transiting material has broken into smaller clumps due to a close passage to its parent star at pericentre. The pericentre and apocentre radii respectively are given by
\begin{equation}
r_{per} = a(1-e)
\end{equation}
\noindent and
\begin{equation}
r_{apo} = a(1+e) .
\end{equation}

\begin{figure}
\includegraphics[width=0.48\textwidth]{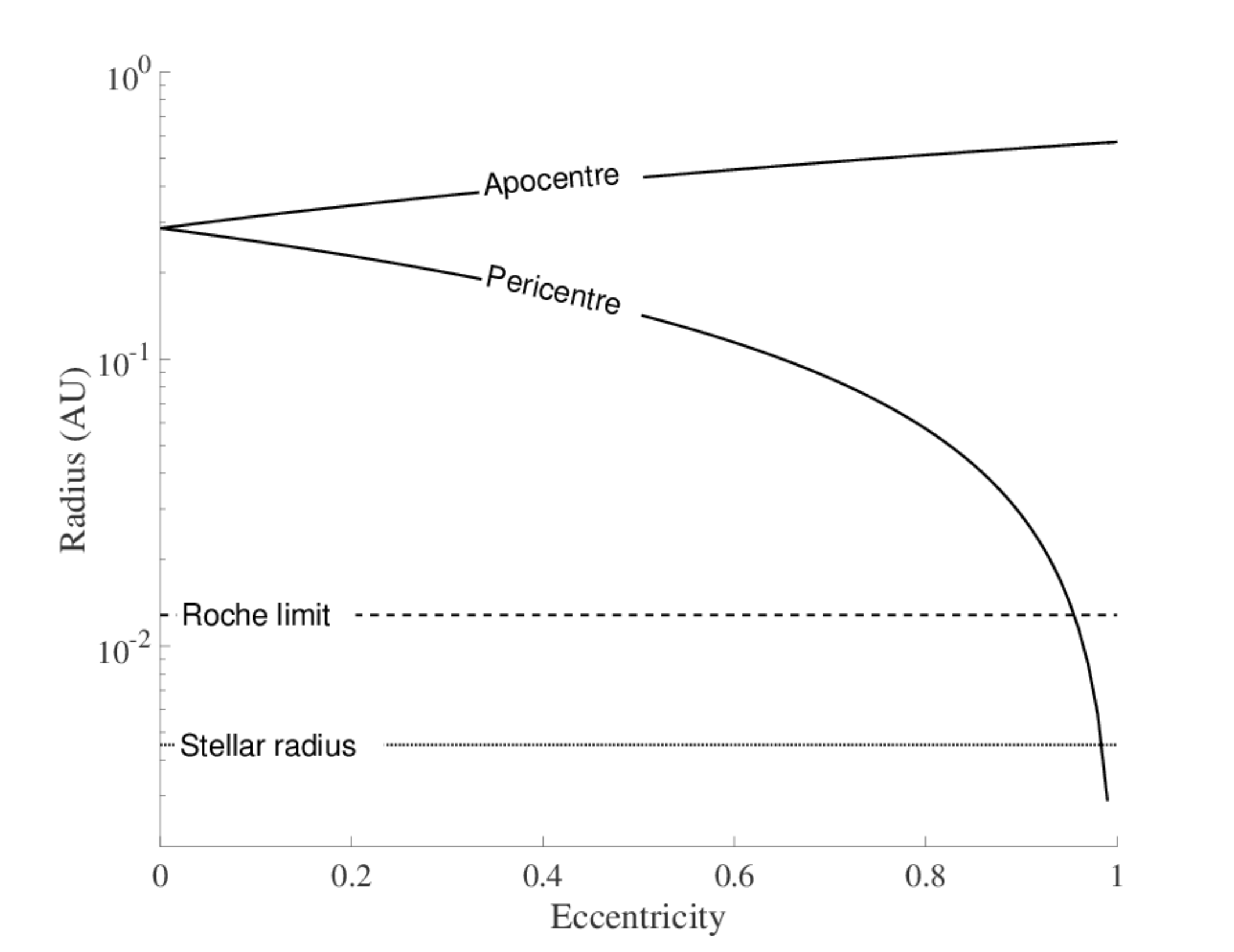}
\caption{Pericentre and apocentre radii for varying eccentricities adopting the observed minimum semi-major axis of $a=0.2855$ AU. The dashed line marks the Roche radius below which a large cometary-like body with density $\rho=0.5$ gm/cm$^3$ would break-up. The dotted line marks the stellar radius, below which orbiting clumps would crash into the star. For this scenario the eccentricity can be constrained to be above 0.95.}
\label{fig:orbRad}
\end{figure}

\noindent Fig. \ref{fig:orbRad} shows these radii as a function of
eccentricity adopting the same minimum semi-major axis as used in Fig.
\ref{fig:orbVel}. The obtained radii in this case can be thought of as
lower limits, since increasing the semi-major axis will result in larger
pericentre and apocentre radii. The dashed line in Fig. \ref{fig:orbRad}
shows the Roche radius below which a cometary-like object with a density
of 0.5 g/cm$^3$ would break-up. The
eccentricity would have to be very large ($e>0.95$) for this case, and might resemble
the Comet Shoemaker-Levy 9 event that broke apart and collided with
Jupiter in July 1994 (see e.g. \citealt{chodas96}).

\subsubsection{Mass constraint of transiting clumps}

We can place a lower constraint on the mass of the transiting material
(independent of eccentricity) by assuming that the observed individual
dips are caused by a cluster of objects, each of a different mass and
size. The mass of each clump is defined by

\begin{equation}
M = m_{g} \rho V ,
\label{eq:mass}
\end{equation}

\noindent where $m_{g}$ is the average grain mass, $\rho$ the object density and $V$ its volume. In turn the volume can be estimated by

\begin{equation}
V = \pi R_{c}^{2} \Delta R ,
\label{eq:vol}
\end{equation}

\noindent where $\Delta R$ is the clump depth along the line of sight. We can obtain an estimate
on $R_c$ from the observed dip depths, but it is non-trivial to obtain
$\Delta R$. However, from the definition of optical depth ($\tau$) we
find that

\begin{equation}
\tau = \rho \sigma \Delta R ,
\label{eq:tau}
\end{equation}

\noindent where $\sigma$ is the cross-section of the particles causing
the obscuration. Through some algebraic manipulation we can substitute
Eq. \ref{eq:tau} and Eq. \ref{eq:vol} into Eq. \ref{eq:mass} to obtain

\begin{equation}
M = \frac{m_g \pi R_{c}^2 \tau}{\sigma} ,
\label{eq:mass2}
\end{equation}

\noindent and note that both the density and clump depth cancel out. Eq. \ref{eq:mass2} is
useful in obtaining a lower mass limit on the transiting material,
since we know the minimum clump radii from the dip depths and also that
$\tau \geq 1$. The only unconstrained parameters are the cross-section
$\sigma$ and the grain mass $m_g$, which we fix to $10^{-8}$ cm$^2$ and $10^{-14}$ grams respectivley, typical for dust grain sizes of $\approx 0.1 \mu$m. These values should be regarded as lower limits from derived dust grain mass distributions (e.g. \citealt{li98}).

\begin{figure*}
\includegraphics[width=1\textwidth]{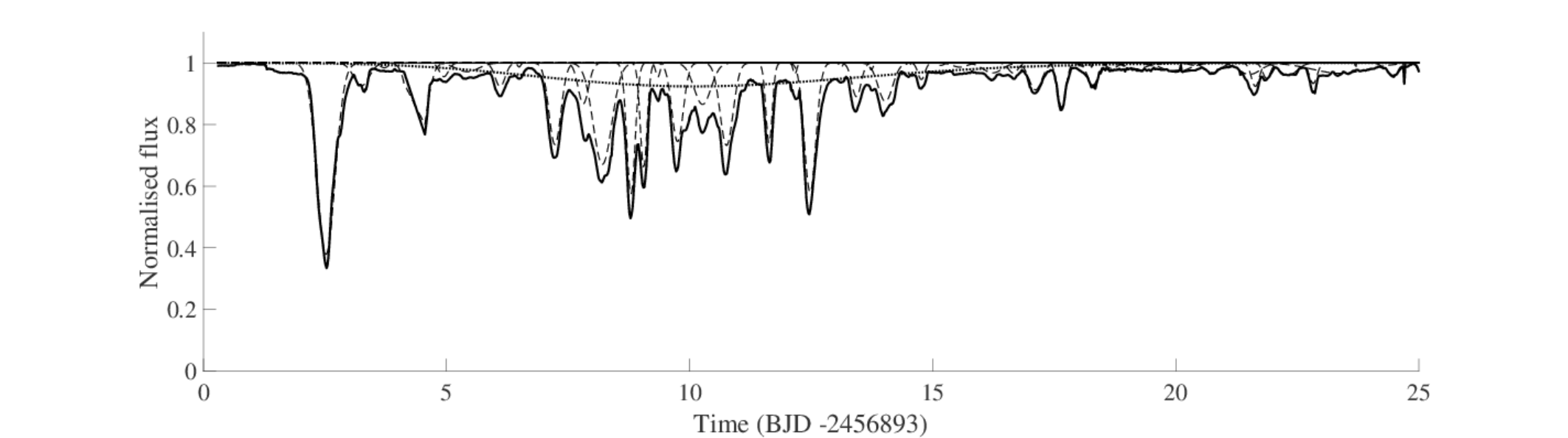}
\caption{Normalised \EPIC\ light curve (solid line, with the 3.646-day periodicity removed) decomposed into 53 gaussians (dashed lines) and one Poissonian component (dotted line).}
\label{fig:fit}
\end{figure*}

To obtain the number and radius of the transiting clumps, we perform a
multi-gaussian fit to the \EPIC\ light curve (after removing the
3.646-day periodicity) shown in Fig. \ref{fig:LC-Spin}. Using a local
peak finding algorithm (\textsc{findpeak} implemented in
\textsc{MATLAB}) we identify 53 dips. Fig. \ref{fig:fit} shows the
first 25 days of observations fitted with 53 gaussians of differing widths
and amplitudes and one broad Poisson function covering the timespan of
the smaller dips. We find that the Poisson
component is necessary to obtain a good fit to the data, but
that the goodness-of-fit is not sensitive to the exact number of
gaussians used. For each gaussian component we obtain the related
$R_c$ and estimate the mass of each individual component using Eq.
\ref{eq:mass2}. The sum of all gaussians then yields a reasonable lower
mass limit for the whole clump of $M_c=7\times10^{17}$ grams
($\approx3.2$ times the mass of Halley's Comet).

\section{Discussion} \label{sec:discussion}

The large-amplitude dipping events observed in the \K2\ lightcurve of
\EPIC\ superficially resembles other systems recently discovered with \K2.
Most notably, \KIC\ (\citealt{boyajian15}) was observed to have similar dip durations to those observed in \EPIC.
However, aside from this, \KIC\ and \EPIC\ differ in
many other respects. Firstly, \EPIC\ shows much deeper dips than \KIC. Furthermore, the dipping patterns observed in \EPIC\ are
clustered in time, with an initial deep dipping event being followed by smaller ones
for $\approx 25$ days before returning to the normal, presumably
quiescent state of the star. The dips observed in \KIC\ are
not clustered, but are spread out over a period of several years. More importantly
however, \KIC\ displays an ordinary F-type star spectrum
(\citealt{boyajian15}) showing no accretion signatures, whilst \EPIC\ is
a YSO still surrounded by a disk, based on both the ALMA image shown in
this work (Fig. \ref{fig:alma}), its spectral features (e.g., presence
of lithium absorption line and H$\alpha$ emission line,
\citealt{preibisch02}), and its membership in the Scorpius-Centaurus OB
association (\citealt{ScoCen}).

More recently \cite{ansdell16b} have discovered 3 objects which more
resemble \EPIC\ when compared to \KIC. These also belong to the
Upper-Scorpius association and are YSOs. Similar to what has been
presented here for \EPIC, \cite{ansdell16b} have resolved the disks with
ALMA, and demonstrated that large-amplitude dipping events are not only
observed in edge-on systems. Given that both the objects presented by
\cite{ansdell16b} and \EPIC\ belong to the same star-forming
association, and all show large dipping events, one might consider them
to be part of the same class of dipping systems. However, the peculiarly
clustered dipping structure displayed by \EPIC\ distinguishes it
from previously reported YSO dipping systems. The only exception might be EPIC 204530046 (presented in \citealt{bodman16}), although we discussed in Section \ref{sec:disk}, why we think \EPIC\ is qualitatively different.

Given the large range of inclination angles inferred from the resolved ALMA
images of \EPIC\ and the sample of \cite{ansdell16b}, it is unlikely
that the observed dips are related to the outer-edges of the
proto-stellar disk. It is however possible that an inclined and variable
inner dust disk could cause some of the observed dips (see e.g. HD
142527, \citealt{marino15}). This is particularly relevant for some of
the systems discussed in \cite{ansdell16b} and \cite{bodman16}, where the dipping events are
observed to persist throughout the full $\approx$ 3 months of \K2\
observations and some display quasi-periodic dips that repeat on the period of the stellar co-rotation radius. If a similar mechanism were responsible for the
observed dips in \EPIC, the inclined inner disk would have to be
transient on a relatively short timescale of a few weeks, making this
interpretation also unlikely. Furthermore we find no relation between the repeating dip patterns in \EPIC\ with the stellar rotation. Thus both the transient nature of the observed dips and the fact that they do not repeat (quasi-)periodically makes \EPIC\ stand out even more from previously observed YSO dippers.

In Section \ref{sec:comets} we explored the possibility that the
observed dips in \EPIC\ are caused by transiting circumstellar clumps.
We showed how circular orbits for this interpretation are most likely
ruled out and that highly eccentric orbits are consistent with the
observations. If transiting cometary-like bodies are responsible for
the observed dips, the events are most likely occurring close to
periastron passage. If the dips are the result of a
previous disruptive event of a larger body, we can further say that the
eccentricity needs to be larger than 0.95 for previous pericentre
passages to be within the Roche radius of the parent star. Similar
disruptive events have been witnessed in our Solar System (e.g. Comet
Shoemaker-Levy 9), but have never been previously witnessed around other
stars. Thus \EPIC\ could constitute the first system where a
planetesimal-sized body has been witnessed to be tidally disrupted by
the parent star upon a close encounter. This would be then a direct
evidence of the presence of km-sized bodies in a protoplanetary disk, a
crucial step towards planet formation.

Given there is no complete explanation for the mysterious behaviour of
\EPIC, more observations and modelling of this system are required to
fully explain the clustered large-amplitude dipping events.
Continuous photometric monitoring of this system for subsequent dipping
events will determine whether this behaviour is periodic or not.
Given the dynamics of
break-up orbits, we would expect the cometary-like bodies to only
survive a few orbits before hitting the parent star. It is thus
important to monitor \EPIC\ before such an event occurs.

\section{Conclusion} \label{sec:conclusion}

We have presented the \K2\ light curve of the disk-bearing young-stellar
object \EPIC, together with a resolved ALMA image constraining its disk
inclination to 57 $\pm$ 9 degrees. The \K2\ light curve displays
prominent, large-amplitude, dips during the first $\approx25$ days
of observations out of the 78.8 day \K2\ observing campaign. Although
difficult to establish their true physical origin, we have discussed the
observed dips in terms of a warped inner disk transiting circumstellar clumps in circular orbits, and cometary-like debris in an eccentric orbit.

It is clear that further observations of \EPIC\ and other YSO dippers
will be required in the future, both photometric and spectroscopic, in
order to establish their true origin. In particular it is important to
determine whether the observed dips in the \K2\ light curve of \EPIC\
are observed again, in which case infer their recurrence timescale and
spectroscopic properties. In this respect we point out the possibility of \K2\ to re-observe part of the Scorpius-Centaurus OB association in 2017 during the planned Campaign 15.

\section*{Acknowledgements}

We gratefully thank the anonymous referee for providing useful and
insightful comments which have improved this manuscript. S.S.
acknowledges funding from the Alexander von Humboldt Foundation. C.F.M
acknowledges ESA research fellowship funding. This research has made use
of NASA's Astrophysics Data System Bibliographic Services. Additionally
this work acknowledges the use of the astronomy \& astrophysics package
for Matlab (\citealt{matlab}). This paper includes data collected by the
Kepler mission. Funding for the Kepler mission is provided by the NASA
Science Mission directorate. Some of the data presented in this paper
were obtained from the Mikulski Archive for Space Telescopes (MAST).
STScI is operated by the Association of Universities for Research in
Astronomy, Inc., under NASA contract NAS5-26555. Support for MAST for
non-HST data is provided by the NASA Office of Space Science via grant
NNX13AC07G and by other grants and contracts. This paper additionally
makes use of the following ALMA data: ADS/JAO.ALMA$\#2013$.1.00395.S.
ALMA is a partnership of ESO (representing its member states), NSF (USA)
and NINS (Japan), together with NRC (Canada) and NSC and ASIAA (Taiwan),
in cooperation with the Republic of Chile. The Joint ALMA Observatory is
operated by ESO, AUI/NRAO and NAOJ.  This material is based upon work
supported by the National Science Foundation Graduate Research
Fellowship under Grant No. DGE-1144469 

%%%%%%%%%%%%%%%%%%%%%%%%%%%%%%%%%%%%%%%%%%%%%%%%%%



\bibliographystyle{mnras}
\bibliography{epicPaperV2} % if your bibtex file is called example.bib

%\begin{thebibliography}{99}
%\end{thebibliography}

%%%%%%%%%%%%%%%%%%%%%%%%%%%%%%%%%%%%%%%%%%%%%%%%%%

%%%%%%%%%%%%%%%%% APPENDICES %%%%%%%%%%%%%%%%%%%%%

%\appendix



%%%%%%%%%%%%%%%%%%%%%%%%%%%%%%%%%%%%%%%%%%%%%%%%%%

% Don't change these lines
\bsp	% typesetting comment
\label{lastpage}
\end{document}